The impact of a carbon nanotube on the cholesterol domain localized on a protein surface

Zygmunt Gburski [*], Krzysztof Górny, Przemysław Raczyński

*Institute of Physics, University of Silesia, Uniwersytecka 4, 40 – 007, Katowice, Poland*

**ABSTRACT**

The influence of a single walled carbon nanotube on the structure of a cholesterol cluster (domain) developed over the surface of the endothelial protein 1LQV has been investigated using the classical molecular dynamics (MD) simulation technique. We have observed a substantial impact of carbon nanotube on the arrangement of the cholesterol domain. The carbon nanotube can drag out cholesterol molecules, remarkable reducing the volume of the domain settled down on the protein.

## 1. Introduction

Cholesterol is one of the most important lipid species in human cells, it is well known to modulate the physical properties of biomembranes. Cholesterol also circulates with the blood stream as a component of lipoproteins and can be found in the lymphatic fluid in the human body. There is a large literature on the role of cholesterol in biosystems [1]. Although cholesterol is essential for the functioning of cell membranes, its excess may prove unhealthy. For example, this starts the development of domains, later foam cells, subsequently leading to formation of plaque deposition in blood vessels [2]. The search for new methods for removing of excess cholesterol molecules, precursors of plaque deposition in an early phase of atherosclerosis disease, is a vital subject of molecular medicine and our simulations are related to this issue. In this context, we have chosen a carbon nanotube since it is known to be hydrophobic. This is an important property when it comes to intervention in the biosystem, where water is the inherent component. The study of the influence of carbon nanotube on cholesterol is in its infancy, limited so far to the very simple nanosystems where only nanotube and cholesterol molecules appear [3, 4]. In this work we made a step towards the investigation of the impact of carbon nanotube on a more realistic, albeit more complicated biosystem fragment. Particularly, we made a reconnaissance study, *via* computer simulation, of the influence of the carbon nanotube on the dynamics of cholesterol molecules forming a domain around a selected extracellular protein in a water environment. Needless to say, taking into account a level of the complexity of the system studied, only the classical MD technique can be effectively applied, *ab initio* simulations would require an enormous computational resources.

* E- mail address:  zgburski@us.edu.pl

## 2. Simulation details

The molecular dynamics (MD) simulations were performed using the NAMD 2.6 program [5] with the all-atom CHARMM force field [6] in NVT (constant number of particles, constant volume and constant temperature) ensemble at the physiological temperature T = 309 K.

The CHARMM force field includes intramolecular harmonic stretching $V_{bond}$, harmonic bending $V_{angle}$, torsional $V_{dihedral}$ terms:

$$V_{total} = V_{bond} + V_{angle} + V_{dihedral} + V_{vdW} + V_{Coulomb} \quad (1)$$

The last two terms in equation (1) describing the total potential energy $V_{total}$ of the system coming from nonbonding interactions: van der Waals forces modeled with the standard Lenard- Jones 12-6 potential $V_{vdW}$ and electrostatic interactions $V_{Coulomb}$ (see Table 1 for the explicit forms of total energy components).

To ensure sufficient energy conservation, the integration time step was set to $\Delta t=0.5$ fs for all simulation runs. The standard NAMD integrator (Brünger-Brooks-Karplus algorithm) was used [4]. As an example, extracellular domain protein 1LQV was chosen (see Protein Data Bank [7]). 1LQV protein appears in the thin layer of cells named endothelium, this layer forms an interface between circulating blood and the rest of the vessel wall. That is why we selected the human protein residing in the innermost layer of a blood artery.

Cholesterol molecules were modeled on the full atomistic level. Fig. 1 presents the model of a cholesterol molecule used in simulation. Atomic charges on cholesterol molecule were taken from [8]. We have chosen to place 21 cholesterol molecules near the surface of the 1LQV protein. Next, to make the environment of the described system similar to that appearing in biological samples, water was added ($15*10^3$ $H_2O$ molecules, TIP3 model [9]). The ensemble consisting of protein, cholesterol and water molecules was equilibrated for $3*10^6$ time steps with periodic boundary conditions in a rectangular simulation box (x = 115 Å, y = 58 Å and z = 78 Å, where x, y and z are the edges of simulation box) [10]. Cutoff distance for all non-bonding interactions was set to 13 Å. After equilibration, the system was simulated for $5*10^6$ time steps (2.5 ns). Trajectories and velocities data were collected every 40 time step.

The aim of our computer experiment was to test whether the carbon nanotube could influence the distribution of cholesterol molecules spread over the protein's surface. First we have done MD simulations for the system composed of 1LQV endothelial protein, cholesterol and water molecules (without nanotube) and the data were collected for reference purpose. Next, an open-ended (uncapped), armchair (10, 10) single wall carbon nanotube (SWCNT) of length 60 Å (960 atoms) was added to the system. The carbon nanotube was also modeled on the atomistic level with internal degrees of freedom allowing oscillations of SWCNT's carbon atoms. As a defalut atom type for carbon nanotube, the CA aromatic carbon CHARMM type was chosen [6,11]. Force field parameters for SWCNT atoms are given in Table 1 (taken from [11]).

Initially, SWCNT was placed at a distance of 14 Å from center of the cholesterol domain and the ensemble was equilibrated during $10^6$ time steps. The configuration of the system with nanotube after equilibration is shown in Fig. 2. A steered molecular dynamics simulation (SMD, see [5]) was performed to bring the SWCNT near the cholesterol lodgment (see Fig. 3). Afterwards, the 2.5 ns production run was executed and MD data were collected over $5*10^6$ time steps. As in the previous case were only 1LQV protein, cholesterol and water appeared, the simulation with nanotube was performed with periodic boundary conditions ( x = 115 Å, y = 58 Å and z = 78 Å) and data were collected every 40 time steps.

## 3. Results and discusion

First, the mean square displacement $\left\langle |\Delta \vec{r}(t)|^2 \right\rangle$ of the center of mass of cholesterol, where $\Delta \vec{r}(t) = \vec{r}(t) - \vec{r}(0)$ and $\vec{r}$ is the position of center of mass of a single molecule, was calculated. From the relation between $\left\langle |\Delta \vec{r}(t)|^2 \right\rangle$ and the translational diffusion coefficient D, $\left\langle |\Delta \vec{r}(t)|^2 \right\rangle \approx 6Dt$ [11] one knows that a nonzero slope of $\left\langle |\Delta \vec{r}(t)|^2 \right\rangle$ is an indicator of mobility

(translational diffusion) of molecules. Fig. 4 shows $\langle |\Delta \vec{r}(t)|^2 \rangle$ in the presence and the absence of the nanotube. It indicates that in both cases the domain is not in the solid state. The diffusion coefficient $D$ of cholesterol, estimated from the linear part of the slope of $\langle |\Delta \vec{r}(t)|^2 \rangle$, in the ensembles studied at T ≈ 309 K is $D = 0.37 * 10^{-6}$ cm$^2$/s (with nanotube) and is $D = 0.39 * 10^{-6}$ cm$^2$/s (without nanotube). Note, that these values of $D$ are related to the short time translational dynamics of cholesterol molecule. If there is no nanotube, the $\langle |\Delta \vec{r}(t)|^2 \rangle$ plot shows saturation around 35 Å$^2$. Therefore, although the cholesterols have some mobility, their displacements are restricted by the interaction with protein surface. One can see that saturation of $\langle |\Delta \vec{r}(t)|^2 \rangle$ vanishes when a SWCNT is placed close to the domain. The cholesterol molecules now can migrate farther, the restriction on their translational dynamics, imposed by the protein surface, is weakened by the competitive interaction with the carbon atoms of the nanotube. Some cholesterols move from the domain and form a layer around the SWCNT. The radial distribution function g(r) of the centers of mass of cholesterols is shown in Fig. 5. In case of system with the carbon nanotube, only the last 0.5 ns trajectories were taken into account in order to examine the final distribution of cholesterols. The presence of the nanotube shifts the average nearest-neighbor distance, from 5.5 Å to 6.6 Å. Naturally, the increase of this distance reflects the fact, that in a final stage some cholesterols still remain within the lodgment and others, pulled out by SWCNT, are spread over nanotube surface.

To visualize even more the migration of cholesterol molecules from the domain we calculated the radial distribution function of cholesterols with respect to the main axis of the carbon nanotube (denoted $g_{CN-chol}$ (r) in Fig. 6). Calculations were performed for both initial and final stage of simulation of system with SWCNT (averaged over 0.2 ns). The radius of nanotube studied is 6.8 Å. Apparently at the final stage some cholesterol molecules approach close to nanotube surface, the first maximum of radial distribution is higher and located nearer the nanotube. To get deeper insight into the migration process we calculated longitudinal and transversal components of the cholesterol displacement (see Fig. 7). The slope of the mean square longitudinal displacement is much lover then the transversal one, so the movement of cholesterols towards the nanotube is faster and overcomes the process of sliding along the nanotube. To show even more directly the ability of SWCNT to remove cholesterol from the domain, another SMD simulation was carried out. In this simulation external forces were applied to pull out SWCNT from the domain (see Fig. 8).

The process of removing cholesterols by carbon nanotube is efficient. The nanotube of 60 Å length has pulled out seventeen of the total number 21 cholesterols, reaching 80 % efficiency. The repeating of this action, using clean nanotube, practically removes remaining cholesterols which survive the first intervention of nanotube. The reported ability of the nanotube to extract the cholesterol lodgment at physiological temperature is quite appealing.

## 4. Conclusions

We have shown that at the physiological temperature T= 309 K the carbon nanotube significantly influences the dynamics of cholesterol molecules which form a layer around the endothelial protein 1LQV. In the presence of the nanotube many cholesterols migrate from the domain (layer) and settle down on the nanotube surface. As a result, the cholesterol domain strongly diminishes. The reported ability of extracting cholesterol domain by carbon nanotube might be related to the quest for future tools of molecular medicine. Naturally, in this context experimental research on the influence of carbon nanotubes on cholesterol lodgments would be of interest. Easier to handle colossal carbon tubes [12] might be a choice in real life experiments.

**Acknowledgment**

NAMD was developed by the Theoretical and Computational Biophysics Group in the Beckman Institute for Advanced Science and Technology at the University of Illinois at Urbana-Champaign.

This work was partly supported by the grant from Polish Ministry of Science and Higher Education.

**Reference:**

[1] T. Róg, M. Pasenkiewicz-Gierula, I. Vattulainen, M. Karttunen, Biochimica et Biophysica Acta 1788 (2009) 97.

[2]  A. J. Lusis, Nature 407 (2000) 233.

[3] P. Raczyński and Z. Gburski,  Biomolecular Engineering, 24(5) 568 (2007).

[4] A. J. Ciani, B. C. Gupta, I. P. Batra, Solid State Communication 147 (2008) 146.

[5] J. C. Phillips, R. Braun, W. Wang, J. Gumbart, E. Tajkhorshid, E. Villa, C. Chipot, L. Skeel, K. Schulten, J Comput Chem. 26 (2005) 1781.

[6] A. D. MacKerell Jr, D. Bashford, M. Bellott, R. L. Dunbrack Jr, J. D. Evanseck, M. J. Field, S. Fischer, J. Gao, H. Guo, S. Ha, D. Joseph- McCarthy, L. Kuchnir, K. Kuczera, F.T.K. Lau, C. Mattos, S. Michnick, T. Ngo, D.T. Nguyen, B. Prodhom, W.E. Reiher, B. Roux, M. Schlenkrich, J.C. Smith, R. Stote, J. Straub, M. Watanabe, J. Wiórkiewicz-Kuczera, D. Yin, M. Karplus, J Phys Chem B 102 (1998) 3586.

[7] Information on http://www.pdb.org

 [8] J. Hénin, C**.** Chipot, Chem. Phys. Letters  425 (2006) 329.

[9] W. L. Jorgensen, J. Chandrasekhar, J. D. Medura, R. W. Impey, M. L. Klein, J. Chem. Phys. 79  (1983) 926.

[10] D. C. Rapaport, *The art of molecular dynamics simulations* (Cambridge University Press, Cambridge 1995).

[11] A. Alexiadis, S. Kassinos, Chem. Rev. 108 (2008) 5014.

[12] H. Peng, D. Chen, Jian-Yu Huang, S. B. Chikkannanavar, J. Hänisch, M. Jain, D. E. Peterson, S. K. Doorn, Y. Lu, Y. T. Zhu, Q. X. Jia, Phys. Rev. Lett. 101 (2008) 145501.

Table 1

CHARMM force field parameters for the interaction between carbon atoms in single-walled carbon nanotube, taken from [6,11]

| $V_{stretch} = K_r(r-r_0)^2$ | |
|---|---|
| $K_r$ [kcal mol$^{-1}$ Å$^{-2}$] | $r_0$ [Å] |
| 305.000 | 1.3750 |
| $V_{bend} = K_\Theta(\Theta-\Theta_0)^2$ | |
| $K_\Theta$ [kcal mol$^{-1}$ rad$^{-2}$] | $\Theta_0$ [degrees] |
| 40.00 | 120.00 |

$$V_{torsional} = \begin{cases} K_\phi(1+\cos(n\phi-\gamma)) & n \neq 0 \\ K_\phi(\phi-\gamma)^2 & n = 0 \end{cases}$$

| $K_\Phi$ [kcal mol$^{-1}$] | n | $\gamma$ [degrees] |
|---|---|---|
| 3.1000 | 2 | 180.00 |

$V_{vdW} = 4\varepsilon\left[\left(\frac{\sigma}{r}\right)^{12} - \left(\frac{\sigma}{r}\right)^6\right]$ with Lorentz-Berthelot mixing rules

| $\varepsilon$ [kcal mol$^{-1}$] | $\sigma/2$ [Å] |
|---|---|
| -0.070000 | 1.992400 |

**Figures captions:**

Fig. 1. Model of cholesterol molecule, for clarity of the picture the hydrogen atoms are not labeled.

Fig. 2. Snapshot of the configuration of 1LQV protein (black), cholesterol molecules (light gray) and carbon nanotube, placed faraway (14 Å) from the domain. Water molecules are drawn in thin gray lines representation.

Fig. 3. Snapshot of the configuration of 1LQV protein (black), cholesterol molecules (light gray) and carbon nanotube pushed nearer to the domain. Water molecules are drawn in thin gray lines representation.

Fig.4 Mean square displacement of the center of mass of cholesterol molecule in the domain in the presence (solid line) and absence (dashed line) of carbon nanotube.

Fig.5. Radial distribution function of the mass centers of cholesterols in the domain.

Fig.6. Radial distribution function of the centers of mass of cholesterols with respect to the main axis (longitudinal) of carbon nanotube, calculated at the beginning and the end of simulation run.

Fig.7. Mean square of the transversal (solid line) and longitudinal (dashed line) components of the displacement of cholesterol molecule with respect to the main axis of carbon nanotube.

Fig.8. Snapshot of the final stage of ensemble after pulling out the carbon nanotube. Carbon nanotube has removed the majority of cholesterols from the domain previously developed over the protein (1LQV) surface.

Fig. 1.

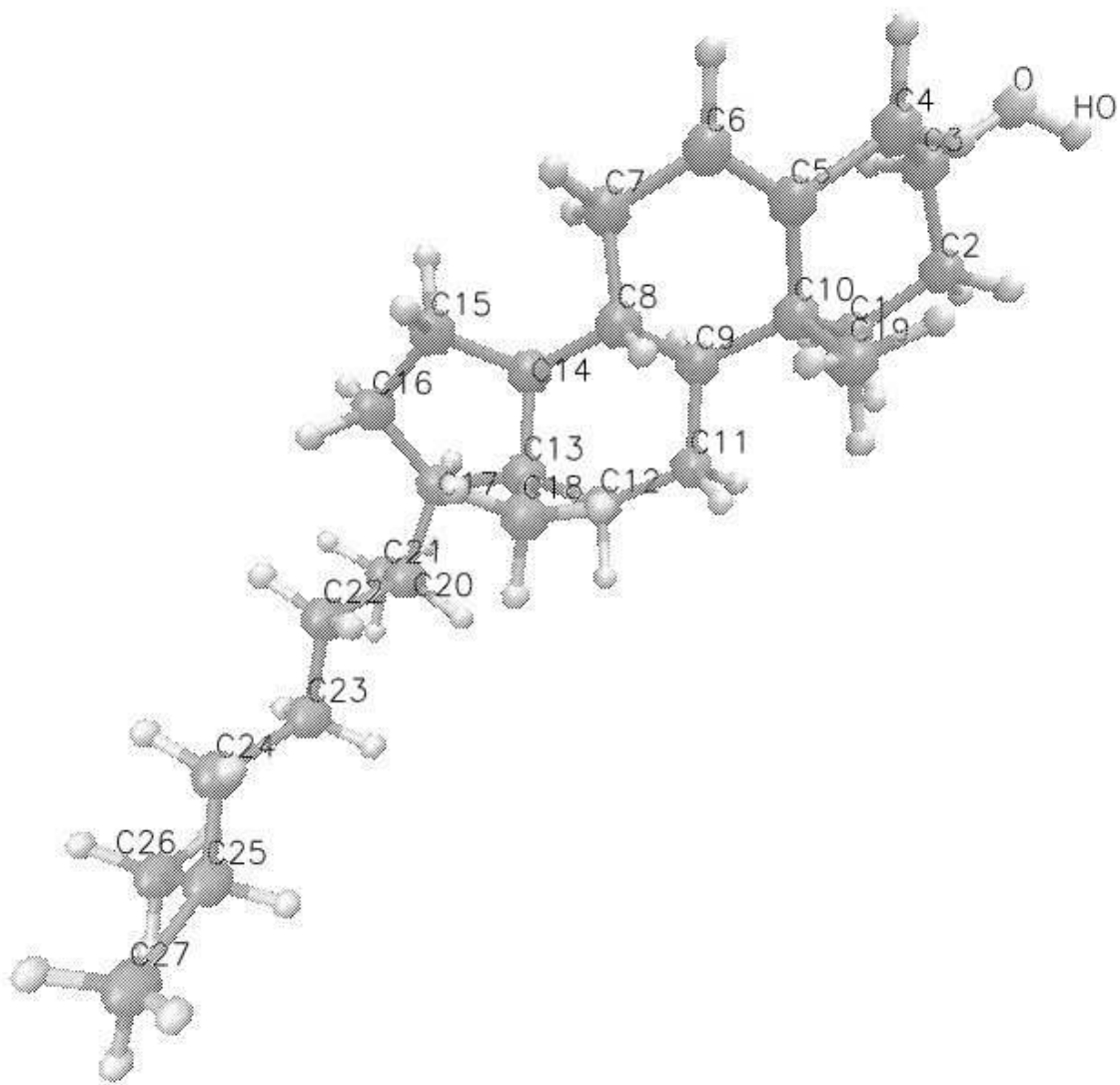

Fig. 2.

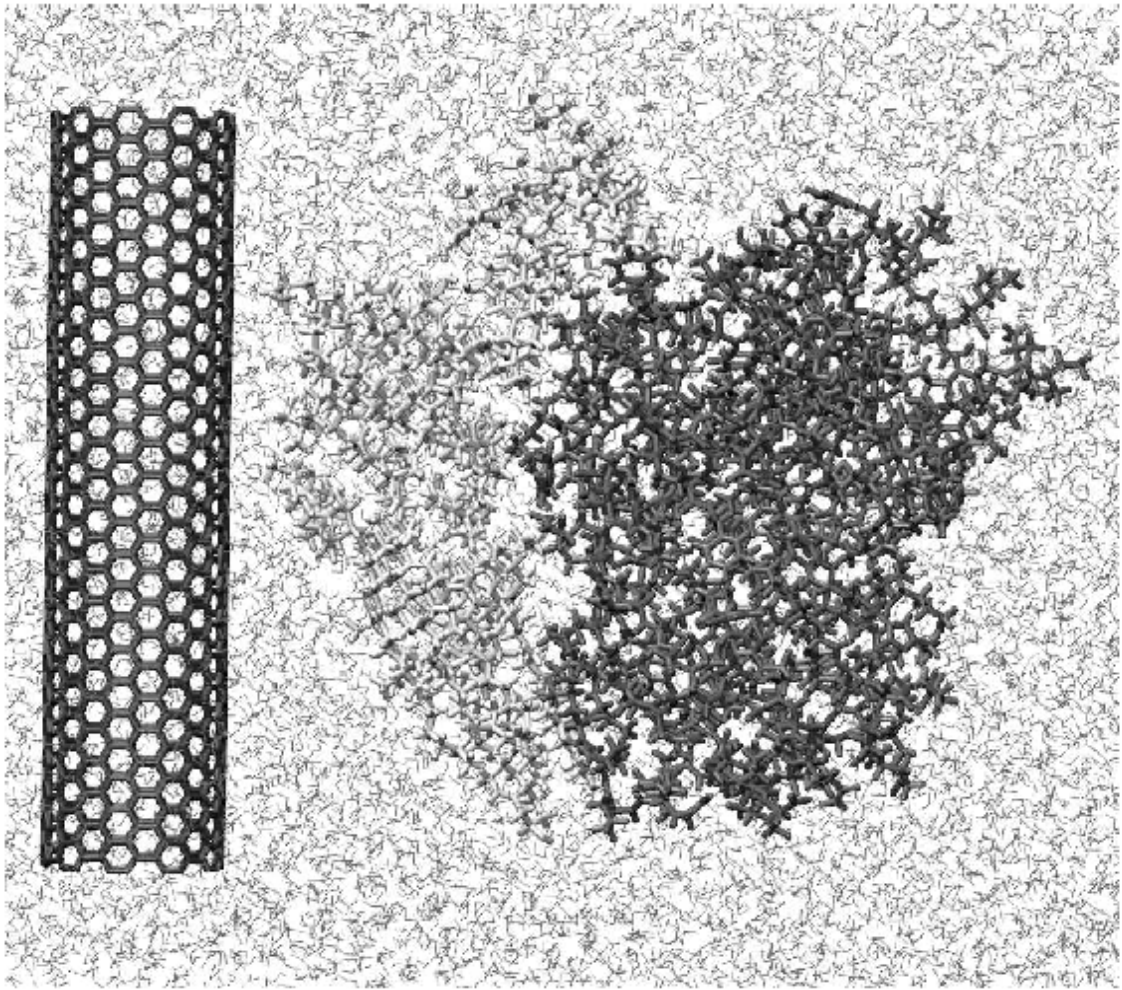

Fig. 3.

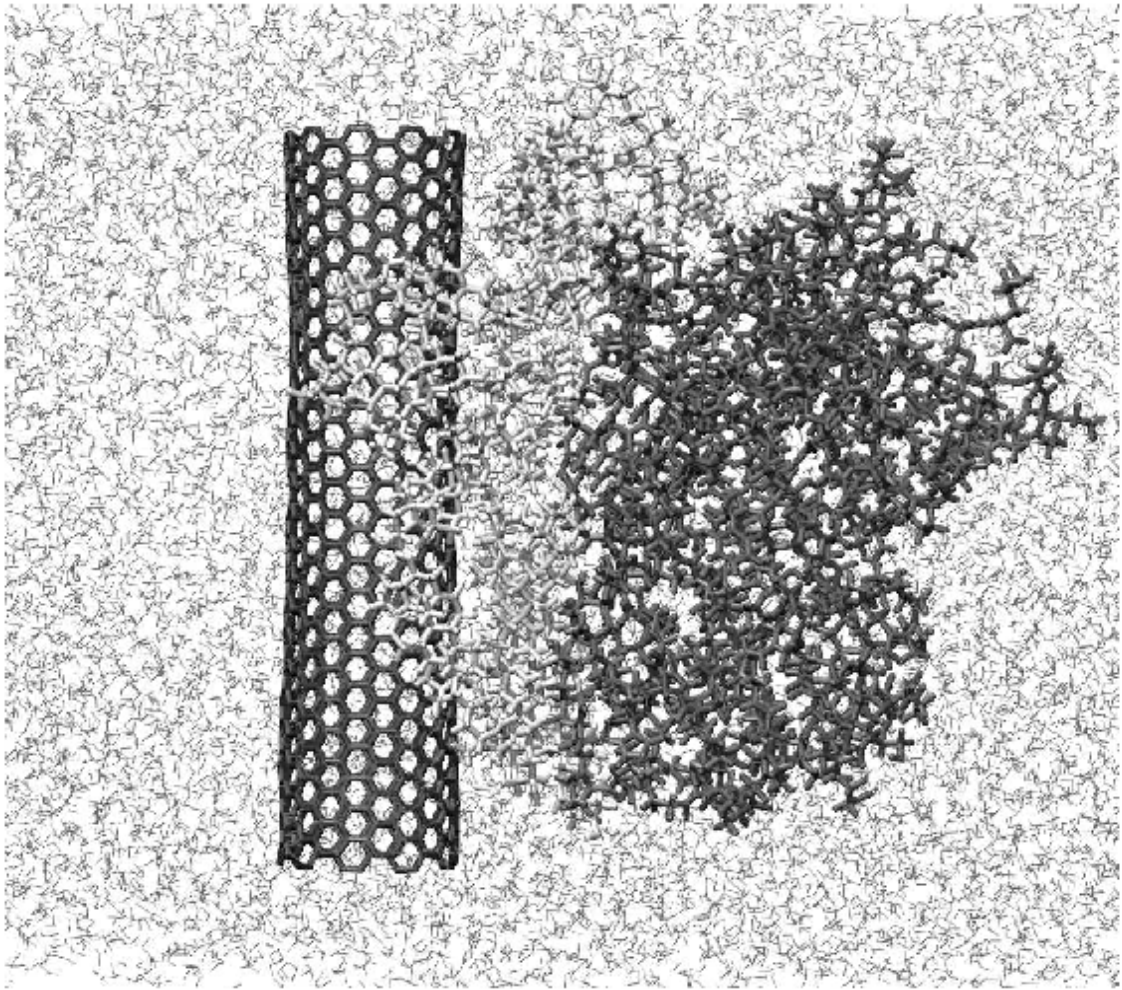

Fig. 4.

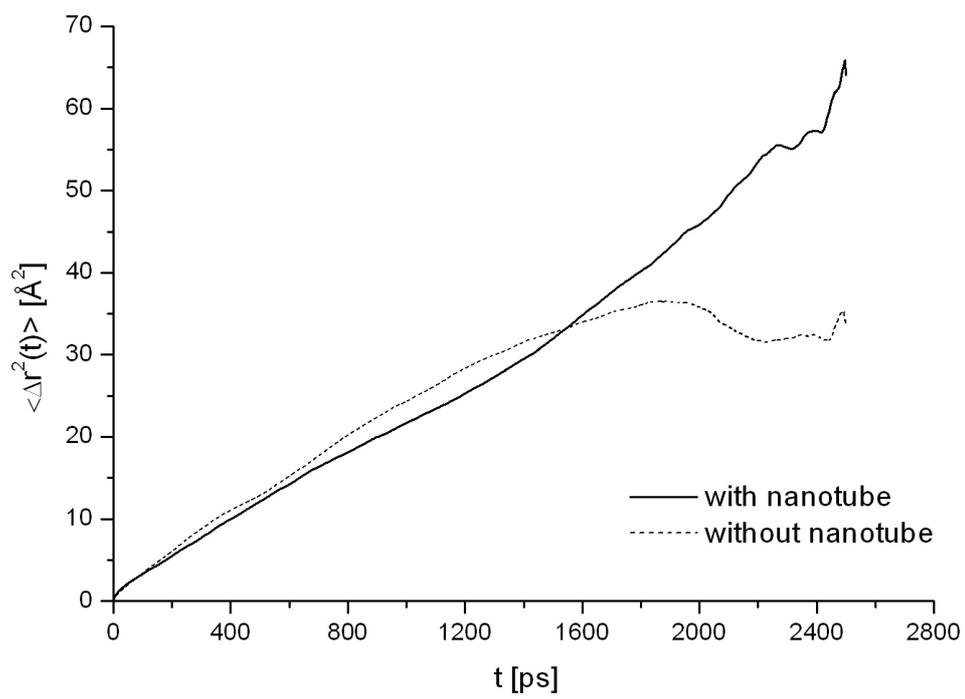

Fig. 5.

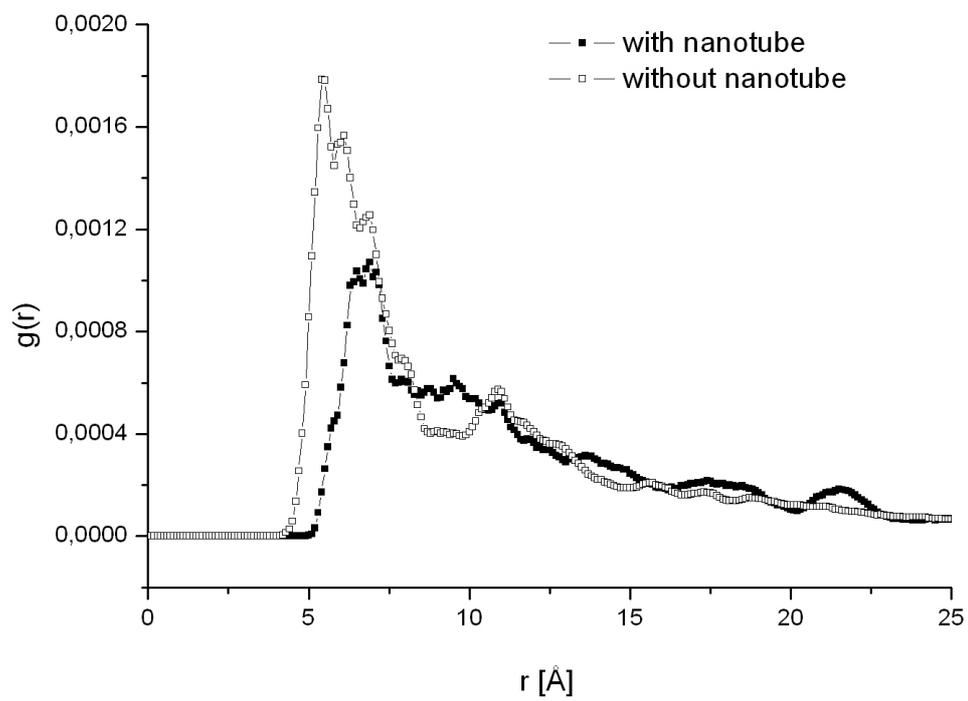

Fig. 6.

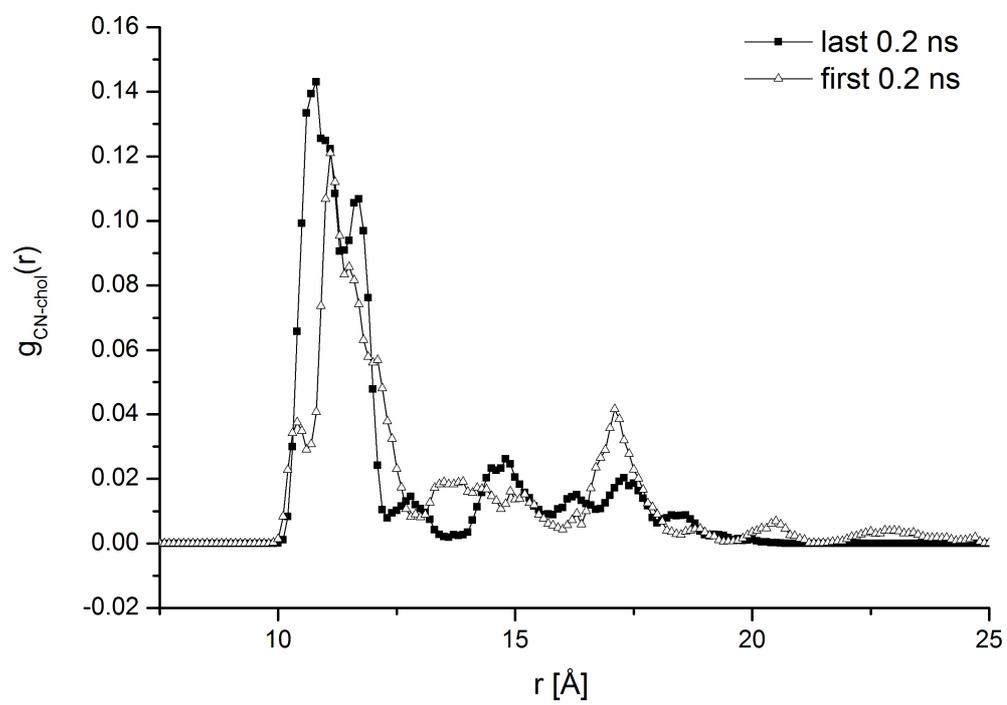

Fig. 7.

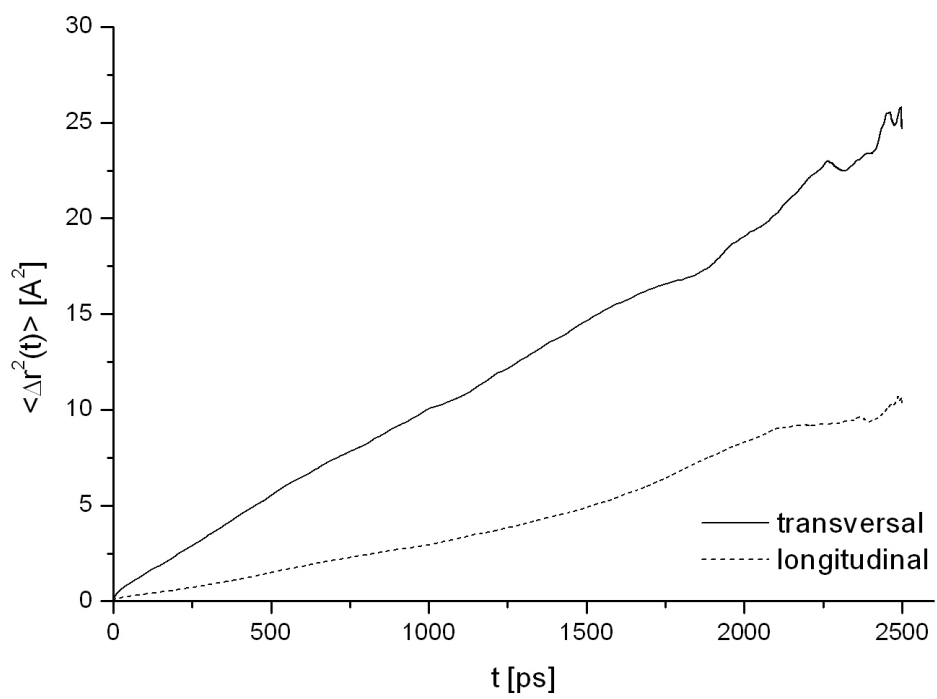

Fig. 8.

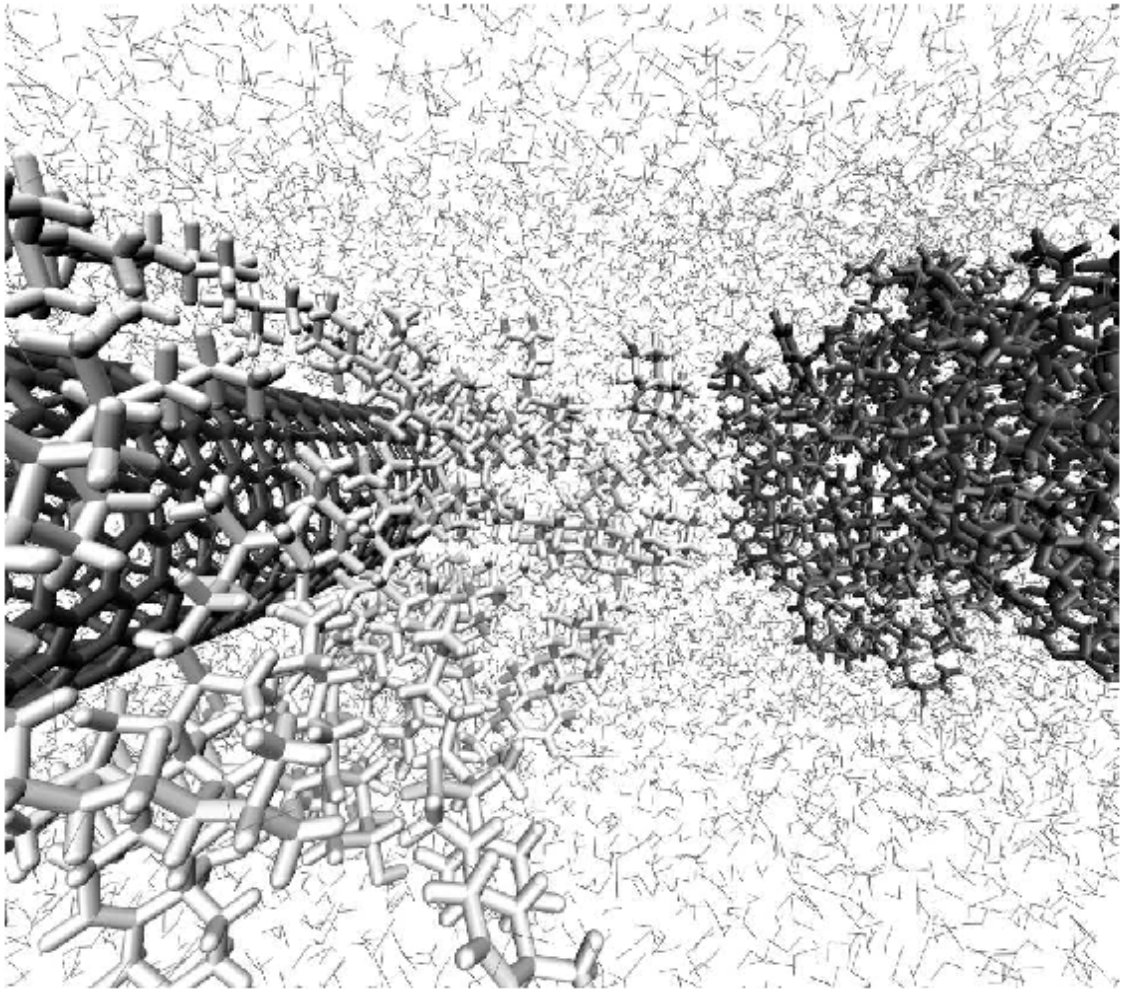